\documentclass[letter]{aa}   
\usepackage{graphicx}
\usepackage{natbib}
\newcommand{\be}{\begin{eqnarray}}         
\newcommand{\ee}{\end{eqnarray}}

\newcommand{\nbodypp}{\textsc{\mbox{nbody6\raise.4ex\hbox{\tiny++}}}}
\newcommand{\COri} {\mbox{$\theta^1{\rm{C}}\:{\rm{Ori}}$}}

\newcommand{\Msun} {\mbox{M$_{\odot}$}}

\begin{document}

\title{Accretion bursts in young stars driven by cluster environment}
\author{S.~Pfalzner, J. Tackenberg, M. Steinhausen} 
\institute{I. Physikalisches Institut, University of Cologne, Z\"ulpicher Str. 77, 50937 Cologne, Germany}
\date{}

\abstract
{The standard picture of accretion is a steady flow of matter from the disc onto the young star - 
a concept which assumes the star-disc system to be completely isolated. However, in a dense cluster environment 
star-disc systems do interact gravitationally.}
{The aim here is to estimate the encounter-induced accretion rate in an ONC-like environment.} 
{Combining simulations of the cluster dynamics with simulations of the effect of encounters on star-disc systems 
we determine the likelihood and degree of encounter-triggered accretion processes.}
{We show that accretion bursts triggered by encounters of star-disc systems are common in young dense clusters 
like the ONC leading in the outburst phase to typical accretion rates of 10$^{-7}$-10$^{-4}$ \Msun/yr. 
Up to a third of stars presently in the Trapezium region accreted at least 1\% of their disc mass via this mechanism 
in the last 1Myr. Accretion of over 6-7\% of the disc material can occur in a single encounter. Despite losing their 
discs quickly, the total percentage of disc matter accreted per star is largest for the massive stars.}
{Supplementing the steady accretion flow there exist episodic periods of high accretion in
dense cluster environments. Due to their high accretion rate these processes should be observable even now in some
of the low-mass stars in the ONC. }

\titlerunning{Accretion bursts in young stars}
\authorrunning{Pfalzner et al.}

\keywords{young clusters - protoplanetary discs - circumstellar matter - ONC}
\maketitle

\section{Introduction}

The theory of accretion has been extensively studied in the context of the standard accretion
disc model \cite{pringle:72,shakura:73,pringle:81}. 
The main idea is that of gas moving inward due to radial turbulent
transport of angular momentum to the outer disc regions. 
Currently the magneto-rotational instability \cite{balbus:91} seems to be able 
to provide the required degree of turbulence in accretion discs.

The disc evolution that follows is not completely understood. 
In the later phases accretion happens exclusively from the disc material, so
linking disc development and accretion directly. The often-used  theoretical models of viscous disc evolution 
\cite{lynden:74,hartmann:98} 
treat the disc as an isolated smooth axisymmetric structure that evolves due to an unspecified 
source of temporally independent, turbulent viscosity.
%
These standard models assume a possibly decreasing with time but relatively {\em steady
accretion} process. Recently  models were suggested where accretion is caused by gravitational instabilities
in the disc \cite{vorobyov:06,boley:06} leading to alternating phases of high and low accretion. However,
for these instabilities to occur the disc has to be relatively massive ($m_d >$ 0.1 $M^*$),
which means it will predominantly occur very early in star formation.

From observations the general picture emerges that the disc fraction is a strong function of age, decreasing 
from $\sim$ 80\% for clusters at 1 Myr down to few discs for
clusters older than 10 Myr \cite{haisch:01,hernandez:07}. 
%
%
The typical accretion rate for a young star is
$\sim$ 10$^{-8}$- 10$^{-10}$\Msun/yr \cite{hillenbrand:95,gullbring:98,haisch:01}. However, the development of
the disc and the accretion rate of individual sources in the same environment can differ considerably
\cite{flaherty:08}.
The accretion rate depends on the stellar age, mass and environment.
As expected from viscous disc models a decrease in accretion rates with stellar age has been
found \cite{hartmann:98, muzerolle:00}. A matter of debate is the 
dependence on the stellar mass of the form $M^\alpha$ 
\cite{calvet:04,sicilia:06,natta:06}, possibly with  $\alpha \sim $ 2.
The effect of the environment possibly manifests itself in that  Herbig Ae stars have on average higher 
accretions rates \cite{lopez:06} than low-mass stars in Taurus and Ophiuchus. 

Here we investigate an accretion process that will be present in dense cluster environments
{\em in addition} to the steady accretion - the passage of a star 
inducing a simultaneous transport of disc matter outwards and inwards via spiral arms, leading to accretion. 
We will show that the consequences are short bursts of high accretion (10$^{-7}$ - 10$^{-4}$ \Msun/yr), 
inevitably occuring in dense clusters and observable due to these high accretion rates.

Although this process has properties in common with accretion induced 
by gravitational instabilities  \cite{vorobyov:06,boley:06} and
in wide binaries 
\citep{bonnell:92},
the cluster dependence is unique to the process treated here.

For encounters to play an important role the cluster has to be dense. A
typical example for such an environment
is the Orion Nebula Cluster(ONC). In an earlier paper \cite{pfalzner:06a}, we demonstrated that 
encounters cause a 3-5\% specific angular momentum loss in the ONC  rising to 10-15\% in the dense inner
Trapezium region. Since specific angular momentum loss is a prerequisite for accretion 
we suggested that in the final star formation stages an additional growth 
mechanism for massive stars exist (cluster-assisted accretion) as
massive stars lose their specific angular momentum to a higher degree than low-mass stars.
Here we study  encounter-induced accretion directly by measuring the
amount of matter kicked into the inner disc areas. Obviously this is not strictly
accretion, but represents a often-used method in circumstances where computational expense does not allow 
to further resolve the inner disc \cite{bate:02,vorobyov:05}. In the following the term ``accreted'' will
be used in this sense but the limitations of this approach should be kept in mind.

  
\section{Method}

We combine simulations of the cluster dynamics performed with 
the Nbody6++ code \cite{spurzem:mnras02} with tree code simulations to study the 
effect of encounters on star-disc systems as described in Pfalzner (2006). 
Here we determine the likelihood and degree of encounter-triggered accretion 
processes. For simplicity all stars of the cluster are assumed to be initially single, the 
effect of gas and the potential of the background molecular cloud OMC~1 are neglected.  
Cluster models were set up with a spherical density distribution $\rho(r)\propto r^{-2}$ and a Maxwell-Boltzmann 
velocity distribution. \COri\ was placed at the cluster centre and assigned a mass of 50 \Msun . For all
other stars the masses were generated randomly according to the mass function 
given by \cite{kroupa:02} in a range $50 \Msun \ge M^* \ge 0.08 \Msun$.
The Nbody6++ code  determines the trajectories of the stars to a high precision \cite[see]{aarseth:book}.
Information of all perturbing events of each stellar disc was recorded during the simulation, i.e. both masses, the relative 
velocity and the eccentricity. More details about the encounter simulations can be found in the
appendix. 
The observed peak density of the Trapezium region of 4.7 $\times 10^4$ stars/pc$^{-3}$ was reproduced 
in our simulations within 20 \%, as was the stellar density distribution of the ONC. 
%
We chose the ONC to be in virial equilibrium. 

The code applied here
cannot describe the disc dynamics  simultaneously with the accretion process. The mass density - 
as the required number of simulation particles - would be too high
close to the star and so would be the required temporal resolution.
In any case the real accretion process will be rather complex. 
The main question is whether perturbations to the inner disc area can be induced
by an encounter. Whether this is done directly by a particle stream or only some related
mechanism is of secondary importance. Therefore we resort to the approach of 
Bate et al (2002) and  Vorobyov \& Basu (2005), by computing the 
amount of matter reaching a sphere around the central star (here 1AU).
We model only gravitational interactions, hydrodynamical effects have been neglected. 
In principle the latter could hinder matter reaching the inner disc areas due to pressure 
gradient. However, here the velocity of the instreaming matter is so
high that this is in most cases not a problem.
%
%
Nevertheless the accretion values attributed here should only be regarded as relative values, 
determining actual accretion rates would require a more sophisticated approach. The present
study only aims at determining in which context encounters are likely to 
trigger accretion processes.

\section{Encounter results}

First we determine how much disc mass is accreted depending on the encounter parameters.
In principle, the star can not only accrete matter from its own disc but 
from the disc of the passing star as well. We find this only happens in very close or
penetrating encounters and so this process is not included in the following.
We performed a parameter study of coplanar, parabolic encounters, which shows that in a single fly-by up to
6-7\% of the disc mass (Fig.~\ref{fig:acc1} and Table~\ref{table:nacc1}) is accreted. 
In general much more mass is lost than accreted, with typical ratios of $m_{accreted}/m_{lost} \sim$ 2-5 \%,
which can rise to $\sim$ 10\% in some cases.
Even fly-bys as distant as 10 $r_d$, where $r_d$ is the disc size (150 AU), 
can induce accretion provided that the mass ratio  
$M^*_2/M^*_1$ of the two interacting stars is high enough, where $M^*_1$ indicates the mass
of the considered star and $M^*_2$ the mass of the passing star.
For $M^*_2/M^*_1 \leq 1$ the amount of accreted matter simply increases the closer 
the encounter, but for very small mass ratios, $M^*_2/M^*_1 <$ 0.1, the accreted 
matter never exceeds 1\% of the disc mass regardless of how close the star passes. 
For $M^*_2/M^*_1 > 1$ the higher mass passing star increasingly captures matter that otherwise 
would have been accreted by the primary.
Note, although the above results were obtained for discs with masses $m_d = 0.001 \Msun$, they are applicable
for $\sim m_d > 0.5 M_1^*$. For a discussion see Appendix.
 



%
%
Typical disc masses are $m_d\sim 0.01 M\sun$ and accretion bursts last normally
$\sim$ 10$^2$ - 10$^3$  yrs. So 5\% accreted disc mass is roughly equivalent to 
an averaged accretion rate of 5 $\times 10^{-7} - 10^{-6}$\Msun/yr. 
However, there will be a wide 
range of induced accretion rates, as the disc mass depends strongly on the stellar 
age and mass.   




So far we considered only prograde encounters, but what happens in retrograde encounters?
As long as the passing star does not penetrate the disc, no matter becomes accreted.
However, encounters closer than 0.5$r_d$ can lead to higher accretion rates in retrograde 
than in prograde encounters. As distant encounters are more common using the accretion rates 
from prograde encounters leads probably to a somewhat overestimated accretion.


Cluster dynamics simulations show that especially in the dense inner regions stars 
often undergo a series of encounters \cite{olczak:apj06}. We simulated consecutive encounters with the same
parameters but using the resulting density distribution as input for the next encounter. 
Each encounter reduces the disc mass and alters the density distribution.
The result is that in subsequent encounters less matter is accreted. To approximate this,
we use the results from single encounters but relative to the reduced disc mass determined 
using Eq.~(4) from Olczak et al. (2006).

%

%
%

In the following we combine above encounter results with cluster simulations described in \cite{olczak:apj06}. 

\begin{figure}
\resizebox{\hsize}{!}{\includegraphics[angle=270]{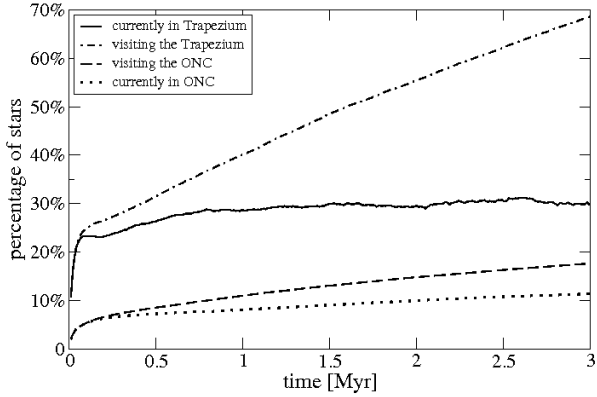}}
\caption{ Percentage of stars currently in Trapezium and ONC region 
with $<$ 1\% of the disc mass accreted as function of time and the same 
for stars that transversed these areas in their past.} 
\label{fig:acc_no_time}
\end{figure}

\section{Cluster-induced accretion}


For ONC conditions
nearly all cluster-induced accretion processes happen within 0.2pc around the cluster center. 
Here the high stellar density leads to a large number of encounters($>$ 500 within 0.03pc 
compared to $<$ 50 outside 0.2 pc). 

After $\sim$ 0.1 Myr about a third of all stars presently in the dense Trapezium region 
have accreted at least 1\% of their disc mass (see Fig.~\ref{fig:acc_no_time}).  
In other words, these stars have shown signs of very high accretion 
($>$ 10$^{-7}$\Msun/yr) due to interacting with neighbouring stars in the past 1-2 Myr. 
The stellar content fluctuates, so looking at stars that have visited the Trapezium 
region in the past, this fraction will increases to $\sim$ 70\% by 3 Myr.
%

\begin{figure}
\resizebox{\hsize}{!}{\includegraphics[angle=-0]{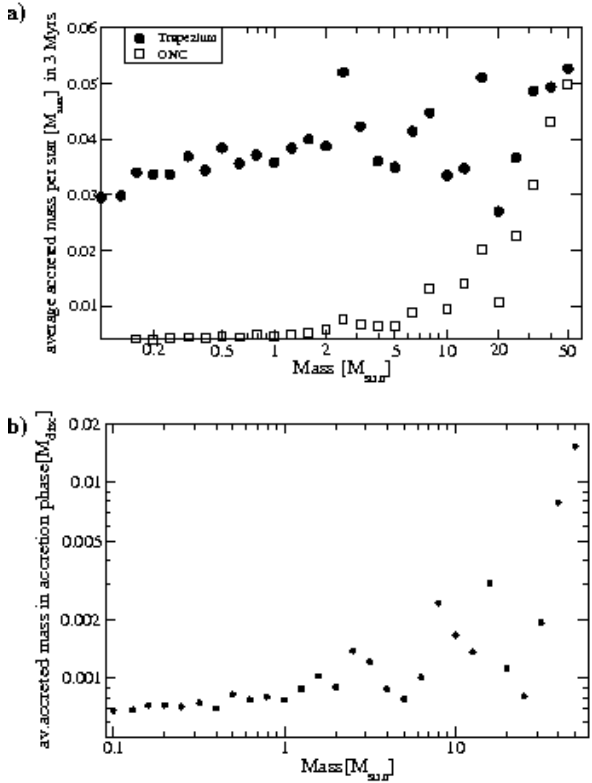}}
\caption{a) Mass accreted per encounter 
in the Trapezium region and the entire ONC  averaged over the first 3 Myrs
and b) mass accreted per star during the accretion phases shown as a function of primary mass
in units of disc masses. }
\label{fig:acc_mass_cluster}
\end{figure}

Fig.~\ref{fig:acc_mass_cluster}a) shows the accreted matter per star as a function
of the primary mass $M_1^*$ during the first 3Myr. In the Trapezium 
region the accreted matter rises by a factor of $\sim$ 1.5 
from low-mass to massive stars. In the entire ONC the dependency
is much stronger as many more low-mass stars never undergo an encounter leading to accretion. 
Here the average accreted mass per star is 10-15 times larger for massive than for 
low-mass stars. 


We find that the accreted mass per accreting encounter 
is nearly constant up to $M_1^* \sim $ 5 \Msun\ and rises for higher masses.
The average number of accreting encounters is also higher for massive stars.
In the Trapezium region a 20 \Msun\ star has on average  more than twice 
the number of accreting encounters  than a 0.1 \Msun\ star; this factor increases to 10 times when
considering the entire ONC. However, the most massive stars lose their discs quickly (typically $\sim$ 5
$\times 10^5$ yrs) so that although they undergo encounters there is no matter left to be accreted.



%
%
Nevertheless, the accreted mass is highest for the most massive stars.
The mass dependence can be most clearly seen looking at the average 
accreted mass in the accretion phase  - meaning the time until the star has completely lost its disc.
Fig.~\ref{fig:acc_mass_cluster}b) shows that the
accreted mass is more than 30 times more for massive stars than low-mass stars.

\begin{figure}
\resizebox{\hsize}{!}{\includegraphics[angle=-90]{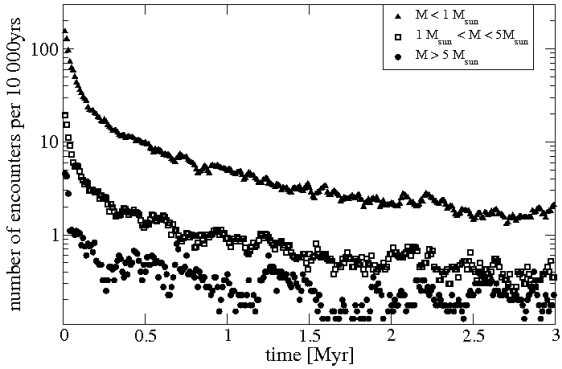}}
\caption{Average number of encounters in any time interval of 10 000 yrs that lead to accretion as a function
of cluster age.}
\label{fig:enc_time_av}
\end{figure}

The number of accreting encounters 
decreases strongly with cluster age, so are there currently any observable encounter-triggered accretion bursts 
in the ONC happening? Although each individual low-mass star is less likely to undergo an accreting encounter than massive stars
due to their relative abundance, one is more likely to observe a low-mass star having such an 
encounter than stars with higher mass (see Fig.~\ref{fig:enc_time_av}). For the low-mass stars the number of encounters 
decreases from several hundred in the first 10$^4$ yrs to  5-10 such events in an timeinterval of 10$^4$ yrs
at the current age of $\sim$ 1 Myr. As most encounter triggered
accretion processes last 10$^3$ - 10$^4$  yrs, this means that one can expect to find currently between 1 and 10 
low-mass stars in the ONC that show high accretion rates due to encounter processes.


\section{Discussion and Conclusion}

There is one observed systems that displays the link between encounter
and accretion rate.
The morphology and kinematics  
\cite{cabrit:06} of RW Aurigae A and B strongly suggest 
tidal stripping of the primary disc by a recent fly-by that
is occurring.
This system displays a high accretion rate of $\sim$ 2-10 $\times 10^{-7}$ \Msun/yr
\cite{basri:89,hartigan:95} while having a particularly low disc mass of $m_d^{c} \sim$ 3 $\times$ 10$^{-4}$ \Msun .
Using these parameters 
our simulations show the likely total accreted mass to be 
0.3 - 4 \% of the pre-encounter mass, 
%
%
equivalent to a maximum induced accretion rate of 1$\cdot 10^{-6}$ - 1.2 $\cdot 10^{-5}$ \Msun/yr assuming 
a pre-encounter disc mass $ m_d^{pre} < 10 \times m_d^{c}$) and  
that most mass is accreted within $\sim$ 10 years 
(although high accretion for several 100 years exist). 
This accretion rate is consistent with the observed accretion rate
of $\sim 2 - 10 \cdot 10^{-7}$ \Msun/yr. So encounter-induced accretion is probably
occuring in this system.

However, RW  Aurigae is not located in a dense cluster, so either the rare case of an encounter 
in a low-density region or a strongly elliptical wide binary is being observed. 
Even in the latter case, the results described in Section 3 are still applicable as the effect 
of a parabolic encounter on the 
induced accretion is very similar to that in a strongly elliptical wide binary.
Episodes of rapid accretion due to tidal disturbances to the disc induced by a binary companion
were suggested by Bonnell and Bastien (1992) in the context of FU Orionis flareups, which are believed
to be powered by episodes of exceptionally rapid accretion (up to 10$^{-4}$ \Msun /yr)
\cite{hartmann:96}. Alternatively, such outbursts could be due to thermal instability 
\cite{bell:94,bell:95}, 
or to gravitational instability in the disc \cite{vorobyov:06}. 
While many of the properties (accretion rate, duration of burst) of the accretion bursts  described  
here agree with those of FUors, the strongest argument against it is the lack of FUors observed in 
dense stellar regions. The arguments for and against a connection of encounter-triggered accretion bursts 
to FU Orionis objects will be discussed in more detail in Pfalzner (2008). 

Accretion bursts might be even more important in the earlier phases
where star-disc systems are still in their formation process \cite{larson:mnras02}.
The simulation of the formation of a small cluster of 
stars by Bate (2002)  clearly shows the formation of many circumstellar discs. The complex dynamics of the system includes
frequent protostellar interactions disrupting the discs by encounters. However new discs soon form 
from gas that continues to fall inwards, so that in this phase protostars may gain much of their mass 
in discrete episodes of encounter-triggered accretion. 

The observed accretion rates in Class I protostars are orders of magnitude smaller 
than those needed to form a star during the lifetime of a Class I object \cite{mckee:07}.
Kenyon et al. (1990) suggested two solutions to this problem - either significant accretion 
continues into the T Tauri stage or  most of the accretion occurs in the 
embedded stage. The former appears to be ruled out because such stars accrete very 
slowly with no significant disc or envelope mass reservoir that 
they can draw on. In the latter case accretion could be episodic, so that the median 
accretion rate is much smaller than the mean.
 
Another fact that can be explained if the accretion occurs in short bursts is that
the observed luminosities of many protostars are lower than is predicted for models
with steady accretion \cite{kenyon:95,calvet:04}. The jet-like Herbig-Haro outflows
probably powered by rapid accretion onto forming stars at early stages of evolution,
are episodic or pulsed,  suggesting that the accretion process is itself episodic
\cite{reipurth:01}. Remarkably the jet sources are frequently (in 85\% of cases) found to have close
stellar companions \cite{larson:mnras02}.


\section{Summary}

We have demonstrated that encounter-induced bursts of accretion are common phenomena in 
the centre of young dense clusters. Successive encounters can lead to over 10\% of the disc mass 
being accreted typically with accretion rates $>$ 10$^{-7}$ \Msun/yr.  
In young dense cluster environments these episodal accretion processes will 
inevitably happen in addition to the steady viscous accretion of disc material.
We showed that accretion-inducing encounters are more likely and lead to higher accretion 
rates for massive stars.



\bibliographystyle{apj}

\begin{thebibliography}{}

\bibitem[Aarseth 2003]{aarseth:book}
Aarseth, S.~J. 2003, {\it Gravitational N-Body Simulations}, CUP, Cambridge, UK.

\bibitem[Alexander \& Armitage 2006]{alexander:06}
Alexander, R.D. \& Armitage,P.J.  2006, {\apj}, 693, L83.

\bibitem[Balbus \& Hawley 1991]{balbus:91}
Balbus, S.A. and Hawley, J.F. 1991, {\apj}, 376, 214.

\bibitem[Basri \& Bertout 1989]{basri:89}
Basri, G. and Bertout, C. 1989, {\apj}, 341, 340.

\bibitem[Bate et al. 2002]{bate:02}
Bate, M.~R. Bonnell, I.~A., Bromm, V. 2002 \mnras 336, 705.

\bibitem[Bell \& Lin 1994]{bell:94}
Bell, K. R., Lin, D.~N.~C. 1996, \apj, 427, 987.

\bibitem[Bell 1999]{bell:95}
Bell, K. R. 1999, \apj, 526, 411.

\bibitem[Boley et al. 2006]{boley:06}
Boley, A.~C., Mejia, A.~C., Durisen, R.H. Cai, K., Pickett, M.~K., 
    D'Alessio, P. 2006, \apj, 651, 517

\bibitem[Bonnell \& Bastien 1992]{bonnell:92}
Bonnell, I. \& Bastien, P. 1992, \apj, 401, L31

\bibitem[Cabrit et al. 2006]{cabrit:06}
Cabrit, S., Pety, J., Pesenti, N., Dougados, C. 2006, \aap, 452, 897.

\bibitem[Calvet et al. 2004]{calvet:04}
Calvet, N., Muzerolle, J., Briceno, C., Hernandez, J., Hartmann, L., Saucedo, J., Gordon, K. 2004, \aj, 128, 1294.


\bibitem[Clarke \& Pringle 2006]{clarke:06}
Clarke, C. \& Pringle, J.E. 2006, \mnras, 370, L10.
	

\bibitem[Dullemond et al. 2006]{dullemond:06}
Dullemond, C. P., Natta, A., Testi, L.  2006, {\apj}, 645, L69.

\bibitem[Flaherty \& Muzerolle 2008]{flaherty:08}
Flaherty, K.M. \& Muzerolle, J. 2008, astro-ph 07121601

\bibitem[Gullbring 1998]{gullbring:98} 
Gullbring, E. Hartmann, L., Briceno, C., Calvet, N. 1998 {\apj} 492, 323. 

\bibitem[Haisch et al 2001]{haisch:01} 
Haisch, K., Lada, E.~A., Lada, C.~J. 2001, {\apj}, 553, l153.

\bibitem[Hartigan et al. 1995]{hartigan:95}
Hartigan, P., Edwards, S., Ghandour, L. 1995, {\apj}, 452, 736.

\bibitem[Hartmann et al. 1998]{hartmann:98}
Hartmann, L.,  1998, {\apj}, 495, 385.

\bibitem[Hartmann et al. 2006]{hartmann:06}
Hartmann, L., D'Allessio, Calvet, Muzerolle 2006, {\apj}, 648, 484.

\bibitem[Hartmann \& Kenyon 1996]{hartmann:96}
    Hartmann, L. \& Kenyon, S.~J. 1996, ARA\&A, 34, 207

\bibitem[Hernandez 2007]{hernandez:07} 
Hernandez, J., Calvet, N., Briceno, C., Hartmann, L., Vivas, A. K., 
Muzerolle, J., Downes, J., Allen, L., Gutermuth, R. 2007, {\apj} 671, 1784.

\bibitem[Hillenbrand  et al. 1995]{hillenbrand:95} 
Hillenbrand, L.~A. et al.  1995, {\aj}, 116, 1816.

\bibitem[Kenyon et al. 1990]{kenyon:90}
Kenyon, S.~J., Hartmann, L. W., Strom, K. M., Strom, S. E.1990, AJ 99, 869. 

\bibitem[Kenyon et al. 1995]{kenyon:95}
Kenyon, S.~J. \& Hartmann, L. 1995 ApJS 101, 117. 


\bibitem[Kroupa 2002]{kroupa:02} 
Kroupa, P. 2002, 295, 82. 

\bibitem[Larson 2002]{larson:mnras02}
Larson, R. B. 2002 {\em MNRAS}, 332, 155.

\bibitem[Lopes et al. 2006]{lopez:06}
Lopes, R.G., Natta, A., Testi, L., Habart, E. 2006, astro-ph/0609032.

\bibitem[Lynden-Bell \& Pringle 1974]{lynden:74}
Lynden-Bell, D. \& Pringle, J.~E. 1974, \mnras, 168, 603.

\bibitem[McCaughren et al. 2002]{mccaughrean:msngr02}
{{McCaughrean}, M.,  {Zinnecker}, H.,  {Andersen}, M., 
    {Meeus}, G.,  {Lodieu}, N.} 2002, {\em The Messenger}, 109, 28.

\bibitem[McKee \& Ostriker 2007]{mckee:07}
McKee, C.~F. \& Ostriker, E.~C. 2007, ARAA 45, 565.

\bibitem[Muzerolle et al. 2000]{muzerolle:00}
Muzerolle, J., Calvet, N., Briceno, C., Hartmann, L., Hillenbrand, L. 2000, {\apj}, 535, L47.

\bibitem[Muzerolle et al. 2005]{muzerolle:05}
Muzerolle, J., Luhmann, K.L., Briceno, C., Hartmann, L., Calvet, N. 2005, \apj, 617, 406.

\bibitem[Natta et al. 2004]{natta:04}
Natta, A. Testi,L., Muzerolle, J., Randich. S., Comeron, F., Persi, P. 2004, \aap, 424, 603.

\bibitem[Natta et al. 2006]{natta:06}
Natta, A., Testi, L., Randich, S. 2006, \aap, 452, 245.

\bibitem[Olczak et al. 2006]{olczak:apj06}
Olczak, C., Pfalzner, S., Spurzem, R. 2006, \apj, 642, 1140.

\bibitem[Ostriker 1994]{ostriker:apj94}
Ostriker, E.C. 1994, \apj, 424, 292.

\bibitem[Pfalzner 2006]{pfalzner:06a}
Pfalzner, S. 2006,  {\apj}, 652, L129.

\bibitem[Pfalzner 2008]{pfalzner:08}
Pfalzner, S. 2008, in preparation.

\bibitem[Pfalzner et al. 2006]{pfalzner:aa06}
Pfalzner, S., Olczak, C., Eckart, A. 2006,  A\&A, 454, 811.

\bibitem[Pringle 1981]{pringle:81}
Pringle, J.~E. 1981 ARAA, 19, 137.

\bibitem[Pringle \& Rees 1972]{pringle:72}
Pringle, J.~E. \& Rees, M.~J. 1972, A\&A, 21,  1.

\bibitem[Reipurth 2001]{reipurth:01}
Reipurth, B., Bally, J. 2001 ARAA 39,  403.


\bibitem[Shakura 1973]{shakura:73}
Shakura, N.~I. 1973, SA,  16, 756.

\bibitem[Sicilia-Aquilar et al. 2006]{sicilia:06}
Sicilia-Aguilar, A., Hartmann, L., Fürész, G., Henning, Th., Dullemond, C., Brandner, W.
2006, AJ 132, 2135 

\bibitem[Spurzem 1999]{spurzem:mnras02} 
Spurzem, R. 1999, Comp. Astroph., 109, 407.

\bibitem[Vorobyov \& Basu 2005]{vorobyov:05}
Vorobyov, E.~I., Basu, S. 2005, \apj 633, L137

\bibitem[Vorobyov \& Basu 2006]{vorobyov:06}
Vorobyov, E.~I., Basu, S. 2006, \apj 650, 956























































































\end{thebibliography}

\Online

\begin{appendix} 

\section{Parameter study} 
 
\subsection{Encounter simulations}

In the encounter simulations themselves the mass of the discs was assumed to be $m_d$=0.001 \Msun\ as is 
typical for T Tauri stars. However, the results are applicable  for higher disc masses as well provided 
self-gravity effects do not play a dominant role. The main effect would be a rearranging of the disc 
material and a deviation from the Keplerian velocities of the disc particles. This probably happens 
for $m_d \sim M_1^*$. For such massive discs the cluster simulation particles would need to represent 
the mass of the star disc system and not just the stars as is currently the case. In this situation the 
encounter rate is likely to rise slightly increasing the number of encounter-induced accretion events 
even more. But this situation would need additional investigations.   

In the encounter simulations the disc size was assumed to be 100 AU. But as the cluster consists of a wide 
spectrum of stellar masses, the simulation results, valid for $M_1^*=1~\mbox{$M_{\odot}$ }$, are generalized 
by scaling the disc radius according to $ r_{\rm {d}} =150 \sqrt{M_1^*[\Msun]},$ 	
which is equivalent to the assumption of a fixed force at the disc boundary.
The quality of the dynamical models were determined by comparing them to observational data 
\cite{mccaughrean:msngr02} at the approximate age of the ONC (1-2Myr). The quantities of 
interest were: number of stars, half-mass radius, number densities, velocity dispersion 
and projected density profile. For more details of the selection process, see \cite{olczak:apj06}.

\begin{figure}[h]
\resizebox{\hsize}{!}{\includegraphics[angle=-90]{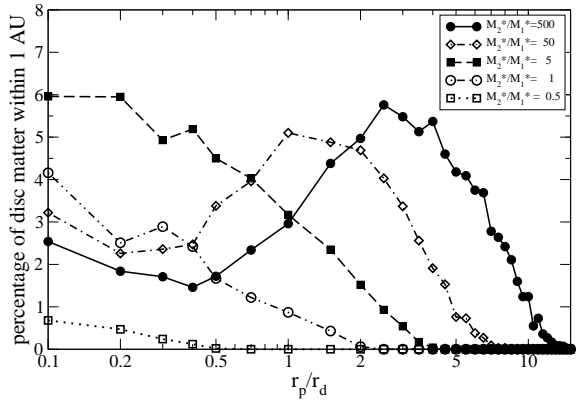}}
\caption{Percentage of disc mass accreted onto star as 
function of the ratio of periastron $r_p$ to the disc radius $r_d$ for different  $M_2^*/M_1^*$ 
ratios in parabolic, prograde coplanar encounters.}
\label{fig:acc1}
\end{figure}

\onecolumn

\begin{table}
\begin{scriptsize}
\begin{center}
\begin{tabular}{r||*{14}{r}}
    & 500.0& 90.0   & 50.0  & 20.0   & 9.0  & 5.0   & 4.0   & 3.0   & 2.0   & 1.5   & 1.0   & 0.5   & 0.3   & 
0.1\\[0.5ex]
\hline
\\[-2ex]
0.1 & 2.54   & 2.86   & 3.22   & 6.94   &  7.83 & 5.96   &  5.85  &   5.10 & 4.33   & 4.22 &   4.16 &   4.26 &   4.87 
&    0.68\\[0.5ex]
0.2 & 1.84   & 1.67   & 2.26   & 4.05   &  5.30 & 5.95   &  6.22  &   5.33 & 5.98   & 4.97 &   2.51 &   3.44 &   2.61 
&    0.47 \\[0.5ex]
0.3 & 1.71   & 2.21   & 2.36   & 2.92   &  4.10 & 4.93   &  4.90  &   5.68 & 5.68   & 5.07 &   2.89 &   2.82 &   1.84 
&    0.24 \\[0.5ex]
0.4 & 1.46   & 2.53   & 2.47   & 3.24   &  4.50 & 5.19   &  4.78  &   4.28 & 4.01   & 3.51 &   2.42 &   2.15 &   1.26 
&    0.12\\[0.5ex]
0.5 & 1.72   & 2.68   & 3.38   & 4.26   &  4.67 & 4.50   &  3.81  &   3.71 & 3.53   & 2.11 &   1.67 &   1.23 &   0.59 
&    0.02\\[0.5ex]
0.7 & 2.34   & 3.65   & 3.96   & 4.77   &  4.91 & 4.03   &  3.74  &   3.14 & 2.24   & 1.83 &   1.22 &   0.54 &   0.22 
&    0 \\[0.5ex]
1.0 & 2.96   & 5.01   & 5.10   & 4.59   &  4.13 & 3.17   &  2.99  &   2.23 & 1.74   & 1.61 &   0.87 &   0.30 &   0.03 
&    0\\[0.5ex]
1.5 & 4.38   & 4.51   & 4.88   & 4.57   &  2.93 & 2.35   &  1.85  &   1.46 & 1.03   & 0.87 &   0.43 &   0.01 &     0 &     
0\\[0.5ex]
2.0 & 4.97   & 5.79   & 4.69   & 3.36   &  2.09 & 1.52   &  1.38  &   1.33 & 0.74   & 0.44 &   0.06 &     0 &     0 &     
0\\[0.5ex]
2.5 & 5.76   & 4.90   & 4.03   & 2.40   &  1.69 & 0.93   &  0.79  &   0.43 & 0.17   & 0.07 &     0 &     0 &     0 &     
0\\[0.5ex]
3.0 & 5.48   & 4.11   & 3.37   & 1.81   &  0.98 & 0.54   &  0.47  &   0.14 & 0.03   & 0.01 &     0 &     0 &     0 &     
0\\[0.5ex]
3.5 & 5.13   & 3.60   & 2.56   & 1.76   &  0.63 & 0.17   &  0.04  &   0.07 &   0   &   0 &     0 &     0 &     0 &     
0\\[0.5ex]
4.0 & 5.37   & 3.02   & 1.91   & 0.79   &  0.14 & 0.04   &  0.01  &   0   &   0   &   0 &     0 &     0  &    0 &   
0.0 \\[0.5ex]
4.5 & 4.60   & 2.31   & 1.53   & 0.56   &  0.11 & 0      &  0     & 0     & 0     &   0 &     0 &     0  &    0 &   0   
\\[0.5ex]
5.0 & 4.18   & 1.76   & 0.76   & 0  8   &  0  2 & 0      &  0     & 0     & 0     &   0 &     0 &     0  &    0 &   0   
\\[0.5ex]
5.5 & 4.09   & 1.56   & 0.73   & 0.11   &  0    & 0      &  0     & 0     & 0     &   0 &     0 &    0   & 0     &   0  
\\[0.5ex]
6.0 & 3.75   & 1.25   & 0.38   & 0  3   &  0    & 0      &  0     & 0     & 0     &   0 &     0   &  0     & 0     & 0  
\\[0.5ex]
6.5 & 3.69   & 0.68   & 0.27   & 0      &  0    & 0      &  0     & 0     & 0     & 0    & 0      & 0    & 0     & 0  
\\[0.5ex]
7.0 & 2.78   & 0.45   & 0  9   & 0      &  0    & 0      &  0     & 0     & 0     & 0    & 0      & 0    & 0  & 0  
\\[0.5ex]
7.5 & 2.64   & 0.15   & 0  3   & 0      &  0    & 0      &  0     & 0    & 0      & 0    & 0      & 0    & 0  & 0  
\\[0.5ex]
8.0 & 2.42   & 0  9   & 0  3   & 0      &  0    & 0      &  0     & 0    & 0      & 0    & 0      & 0    & 0  & 0  
\\[0.5ex]
8.5 & 2.11   & 0  3   & 0      & 0      &  0    & 0      &  0     & 0    & 0      & 0    & 0      & 0    & 0  & 0  
\\[0.5ex]
9.0 & 1.60   & 0  2   & 0      & 0      &  0    & 0      &  0     & 0    & 0      & 0    & 0      & 0    & 0  & 0  
\\[0.5ex]
9.5 & 1.24   & 0  1   & 0      & 0      &  0    & 0      &  0     & 0    & 0      & 0    & 0      & 0    & 0  & 0  
\\[0.5ex]
10. & 1.24   & 0      & 0      & 0      &  0    & 0      &  0     & 0    & 0      & 0    & 0      & 0    & 0  & 0  
\\[0.5ex]
10.5 & 0.55  & 0      & 0      & 0      &  0    & 0      &  0     & 0    & 0      & 0    & 0      & 0    & 0  & 0  
\\[0.5ex]
11.0 & 0.73  & 0      & 0      & 0.     &  0    & 0      &  0     & 0    & 0      & 0    & 0      & 0    & 0  & 0  
\\[0.5ex]
11.5 & 0.36  & 0      & 0      & 0      &  0    & 0      & 0      & 0    & 0      & 0    & 0      & 0    & 0  & 0  
\\[0.5ex]
12.0 & 0.27  & 0      & 0      & 0      &  0    & 0      & 0      & 0    & 0      & 0    & 0      & 0    & 0  & 0  
\\[0.5ex]
12.5 & 0.17  & 0      & 0      & 0      &  0    & 0      & 0      & 0    & 0      & 0    & 0      & 0    & 0  & 0  
\\[0.5ex]
13.0 & 0  9  & 0      & 0      & 0      &  0    & 0      & 0      & 0    & 0      & 0    & 0      & 0    & 0  & 0  
\\[0.5ex]
13.5 & 0  8  & 0      & 0      & 0      &  0    & 0      & 0      & 0    & 0      & 0    & 0      & 0    & 0  & 0  
\\[0.5ex]
14.0 & 0  7  & 0      & 0      & 0      &  0    & 0      & 0      & 0    & 0      & 0    & 0      & 0    & 0  & 0  
\\[0.5ex]
14.5 & 0  2  & 0      & 0      & 0      &  0    & 0      & 0      & 0    & 0      & 0    & 0      & 0    & 0  & 0  
\\[0.5ex]
15.0 & 0  2  & 0      & 0      & 0      &  0    & 0      & 0      & 0    & 0      & 0    & 0      & 0    & 0  & 0  
\\[0.5ex]
15.5 & 0     & 0      & 0      & 0      &  0    & 0      & 0      & 0    & 0      & 0    & 0      & 0    & 0  & 0  
\\[0.5ex]
16.0 & 0     & 0      & 0      & 0      &  0    & 0      & 0      & 0    & 0      & 0    & 0      & 0    & 0  & 0  
\\[0.5ex]
\end{tabular}
\caption{Table of the percentage of particles accreted onto star 1 for all simulated configurations of prograde 
parabolic $(e=1)$ star-disc encounters. The first row contains the relative perturber masses $M_2^*/M_1^*$, the first 
column contains the relative periastra $r_{\rm{p}}/r_{\rm{d}}$. 
\label{table:nacc1}}
\end{center}
\end{scriptsize}
\end{table}

\twocolumn
\end{appendix}

\end{document}